\documentclass[aps,prx,twocolumn,twoside,floatfix,a4,showpacs,superscriptaddress]{revtex4}

\usepackage{amsmath,amssymb,amsfonts}
\usepackage{graphicx}
\usepackage{color}
\usepackage{braket}

\usepackage{ulem}
\begin{document}

\title{A fast fault-tolerant decoder for qubit and qudit surface codes} 
\author{Fern H.E. Watson}
\email{fern.watson10@imperial.ac.uk}
\affiliation{Department of Physics and Astronomy, University College London, Gower Street, London WC1E 6BT, UK.}
\affiliation{Department of Physics, Imperial College London, Prince Consort Road, London SW7 2AZ, UK.}
\author{Hussain Anwar}
\affiliation{Department of Mathematical Sciences, Brunel University, Uxbridge, Middlesex UB8 3PH, UK.}
\author{Dan E. Browne}
\affiliation{Department of Physics and Astronomy, University College London, Gower Street, London WC1E 6BT, UK.}

\begin{abstract}
The surface code is one of the most promising candidates for combating errors in large scale fault-tolerant quantum computation. A fault-tolerant decoder is a vital part of the error correction process---it is the algorithm which computes the operations needed to correct or compensate for the errors according to the measured syndrome, even when the measurement itself is error prone. Previously decoders based on minimum-weight perfect matching have been studied. However, these are not immediately generalizable from qubit to qudit codes.
In this work, we develop a fault-tolerant decoder for the surface code, capable of efficient operation for qubits and qudits of any dimension, generalizing the decoder first introduced by Bravyi and Haah [Phys. Rev. Lett. 111, 200501 (2013)]. We study its performance when both the physical qudits and the syndromes measurements are subject to generalized uncorrelated bit-flip noise (and the higher dimensional equivalent). We show that, with appropriate enhancements to the decoder and a high enough qudit dimension, a threshold at an error rate of more than $8$\% can be achieved.
\end{abstract}


\maketitle

\section{Overview} 

Topological quantum codes built from qubits ($ 2 $-dimensional quantum systems) play a central role in architectures for fault-tolerant quantum computing at the forefront of current research \cite{RHG07,RH07,DFSG09,FMMC12}. The surface code \cite{BK98}, and the related toric code \cite{K03,DKLP02}, are prominent examples of such codes. Compared with other quantum error correcting codes, they posses the key experimental benefit of requiring only local interactions and yet, under realistic noise models, they have been shown to achieve the highest reported fault-tolerant thresholds \cite{WFH11,S14}.

Recent developments have shown that employing $ d $-dimensional quantum systems, or qudits, as the building blocks for fault-tolerant schemes may offer some important advantages. For example, an integral part of many fault-tolerant schemes is the distillation of magic states \cite{BK05}---a procedure necessary to achieve universal computation---where generalization to higher dimensions has resulted in improved distillation thresholds and lower overheads in the number of qudit magic states \cite{ACB12,CAB12,C14}. Moreover, threshold investigations of the qudit toric code with noise-free syndrome measurements have shown that, for a standard independent noise model, the error correction threshold increases significantly with increasing qudit dimension \cite{CP13,ABCB14,WAK14}, although we caution that it is difficult to fairly compare noise rates between systems of different dimension. Although it is more challenging to realize qudit quantum systems experimentally, recent work has demonstrated the ability to coherently control and perform operations in single 16-dimensional atomic systems with high fidelity \cite{SA13,ASM14}, with the implementation of high fidelity multi-qudit interactions still to be achieved.

\begin{figure}
\includegraphics[width=0.35\textwidth]{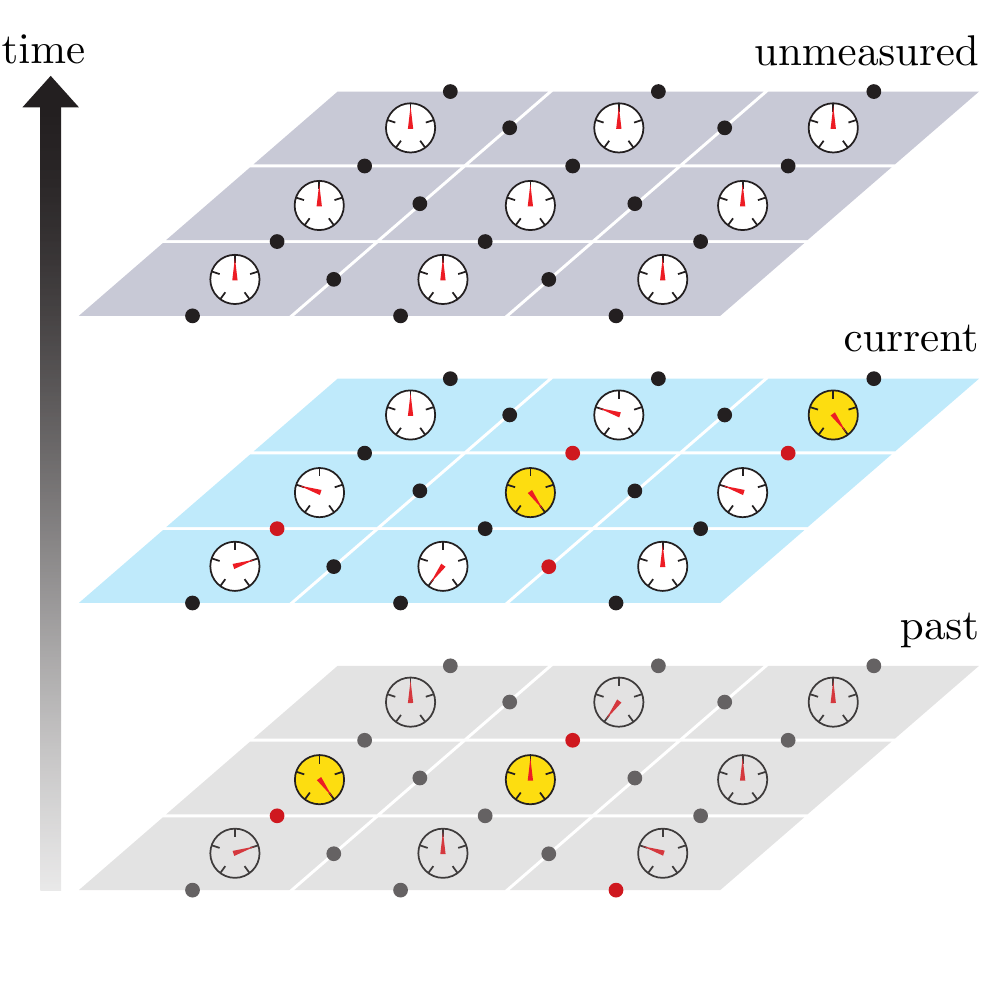} \centering
\caption{(Color online) An illustrative picture of the data structure obtained in order to perform fault-tolerant error correction. Each layer represents a single time step where all the stabilizers are measured (only plaquettes are shown here for clarity---depicted by the meter with multiple outcomes) to obtain the syndrome. The yellow meters (dark gray) represent locations where an error has occurred in the measurement procedure itself. After a specified number of time steps a full 3D history of the syndromes will have been be collected. If operating below threshold the decoder then uses this data to infer a correction operator that returns the code its original state with high probability.}
\label{fig:datastructure}
\end{figure}

A surface code is a stabilizer code with local stabilizer generators. Qudits are associated with the edges of a 2D square lattice. In order to store the encoded information for an arbitrary length of time, active error detection must be performed periodically in order to prevent the errors from accumulating beyond the capability of the code to correct them, see Fig. \ref{fig:datastructure}. In every round of error correction all the stabilizer generators are measured to obtain the \textit{syndrome}. The syndrome is then processed by the \textit{decoder}---the classical algorithm that outputs a correction operator. In a realistic environment both the physical systems and the stabilizer measurements are prone to errors, and hence the decoder must be able to take both of these types of errors into account \cite{DKLP02, WHP03}.

Decoders are often developed for the simpler case where measurement error is neglected. However, there is a well established and elegant method for generalizing measurement-noise-free decoders for topological codes to the fully fault-tolerant setting \cite{DKLP02}. The noisy syndrome measurements are repeated, extending the two-dimensional surface representing the code to a three-dimensional data structure, where time represents an extra dimension. Remarkably, the change from two to three dimensions allows most decoder algorithms developed for noise-free measurements to be applied largely unchanged in this more general setting.

The most widely used decoding algorithm for topological codes remains the minimum-weight perfect matching algorithm (MWPMA). However, this algorithm has a number of disadvantages. For a distance $L$ surface code, with error-free measurements, the run time for a basic implementation of the MWPM algorithm scales with $O(L^6)$, and for with error-prone measurements this run time increases to $O(L^9)$. A more refined fault-tolerant implementation for the qubit surface code scales with $O(L^2)$ \cite{FWH12}, and under certain assumptions a run time complexity that is independent of $L$ can be attained \cite{F15}. Nevertheless, the main disadvantage of MWPMA is that it is not suitable for qudit surface codes with  $d>2$. For these reasons, the development of alternative decoding algorithms is currently a very active research area \cite{CP13, W13, CP14, ABCB14, HWL14, HCEK14, BSV14, HLW14}.

In this work, we introduce a fault-tolerant decoding algorithm which overcomes both of the disadvantages of the MWPMA. The algorithm, which extends the hard-decision renormalization group (HDRG) decoder proposed by Bravyi and Haah \cite{BH13}, has a fast typical run time of $O(L^3)$ and can be applied to qudit surface codes of any dimension $d$.

For a given noise model, the error threshold represents an upper bound on the noise level for which increasing the code distance increases the probability of successful error correction. We denote the threshold for a given qudit dimension by $p^{(d)}_{\mathrm{th}}$. A widely-studied qubit error model (described below in more detail) is the simple \textit{uncorrelated noise} model where $X$ and $Z$ Pauli errors on individual code qubits and bit-flip errors on the syndrome measurement outcomes each occur independently with probability $p$. For this noise model, the optimal threshold for the qubit toric code is known to be $3.3$\% \cite{OAIM04} while the threshold obtained with the MWPMA decoder is $2.9$\% \cite{H04,WHP03,S14}.

The HDRG decoder we study here attains a threshold of $p^{(2)}_{\mathrm{th}}=2.2\%$ for the qubit code and may also be used with qudit surface codes of any dimension. For the qudit generalization of the uncorrelated noise model (introduced below), the decoder achieves a threshold value which increases monotonically with the qudit dimension $d$, until it reaches a saturated value of around $4.2$\%. 

We show that this saturating behaviour is due to a syndrome percolation effect which upper bounds the acheivable threshold. To overcome the percolation threshold we have constructed a procedure executed \textit{before} running the HDRG, which we call the \textit{initialization step} \cite{ABCB14}. The algorithm implemented in this `pre-decoding' step disrupts the syndrome percolation and boosts the threshold to $8.3$\% for sufficiently high qudit dimension. We call the HDRG decoder when augmented with the initialization step the \textit{enhanced}-HDRG decoder.

The structure of this paper is as follows. We start in Sec. \ref{SecQSC} by reviewing the properties of the qudit surface code and fixing our notation. In Sec. \ref{SecNMSM} we give a formal description of the noise model investigated and describe how our numerical simulations were performed. In Sec. \ref{SecHDRG} we present our different variations of the HDRG decoder for the fault-tolerant setting, along with the thresholds we obtain. We conclude in Sec. \ref{SecConclusion}.

\section{The Qudit Surface Code}\label{SecQSC}

The qudit surface code is the natural higher dimensional generalization of the qubit code. This generalization is already present in Kitaev's seminal paper \cite{K03} and has been written about extensively elsewhere \cite{K02, DKLP02,WHP03, BB07, FWH10}. For completeness, however, we shall provide an overview of qudit stabilizer codes and the qudit surface code. 

We express the computational basis for a single qudit as the set of states $\ket{\alpha}$ where $\alpha \in \mathbb{Z}_d$, and where the $d$-element cyclic group $\mathbb{Z}_d=\{0,\dots,d-1\}$ can be conveniently identified with addition over integers modulo $ d $. The conventional single qubit Pauli operators have natural generalizations:
\begin{eqnarray}
X = \sum_{j\in \mathbb{Z}_d} \ket{j \oplus 1}\bra{j}, \hspace{3mm} Z = \sum_{j\in \mathbb{Z}_d} \omega^{j} \ket{j} \bra{j},
\label{eqn:single_Paulis}
\end{eqnarray}
where $\omega = e^{2 \pi i/d}$ and the addition `$\oplus$' is taken to be modulo $d$. Notice that these unitary operators are no longer Hermitian when $ d>2 $, but they posses orthogonal eigenspaces with eigenvalues of the form $ \omega^{j} $, for some $j$. Hence, we can still interpret them as physical observables with measurement outputs labelled by their complex eigenvalues. As a shorthand we will often abbreviate an outcome $ \omega^{j} $ simply by its exponent $ j $.

The qudit Pauli operators obey the commutation relation $X^j Z^k = \omega^{-jk} Z^k X^j$ for arbitrary $j,k \in \mathbb{Z}_d$. They generate the single qudit Pauli group $\mathcal{P}_{d} = \left< X, Z \right>$ up to a global phase. The $n$-qudit Pauli group $\mathcal{P}^{n}_{d}$ is the $n$-fold tensor product of the single qudit Pauli group $\mathcal{P}^{\otimes n}_{d}$. The code space of a stabilizer code is defined as the `$+1$' eigenspace of an abelian subgroup $\mathcal{S} \in \mathcal{P}^{n}_{d}$, such that $\omega^j \openone \not\in \mathcal{S}$ for non-zero $j$. The elements of $\mathcal{S}$ are called the \textit{stabilizers} of the code. A set of generators of $\mathcal{S}$ is identified as the syndrome measurement operators for the code. 

\begin{figure}
\includegraphics[width=0.4\textwidth]{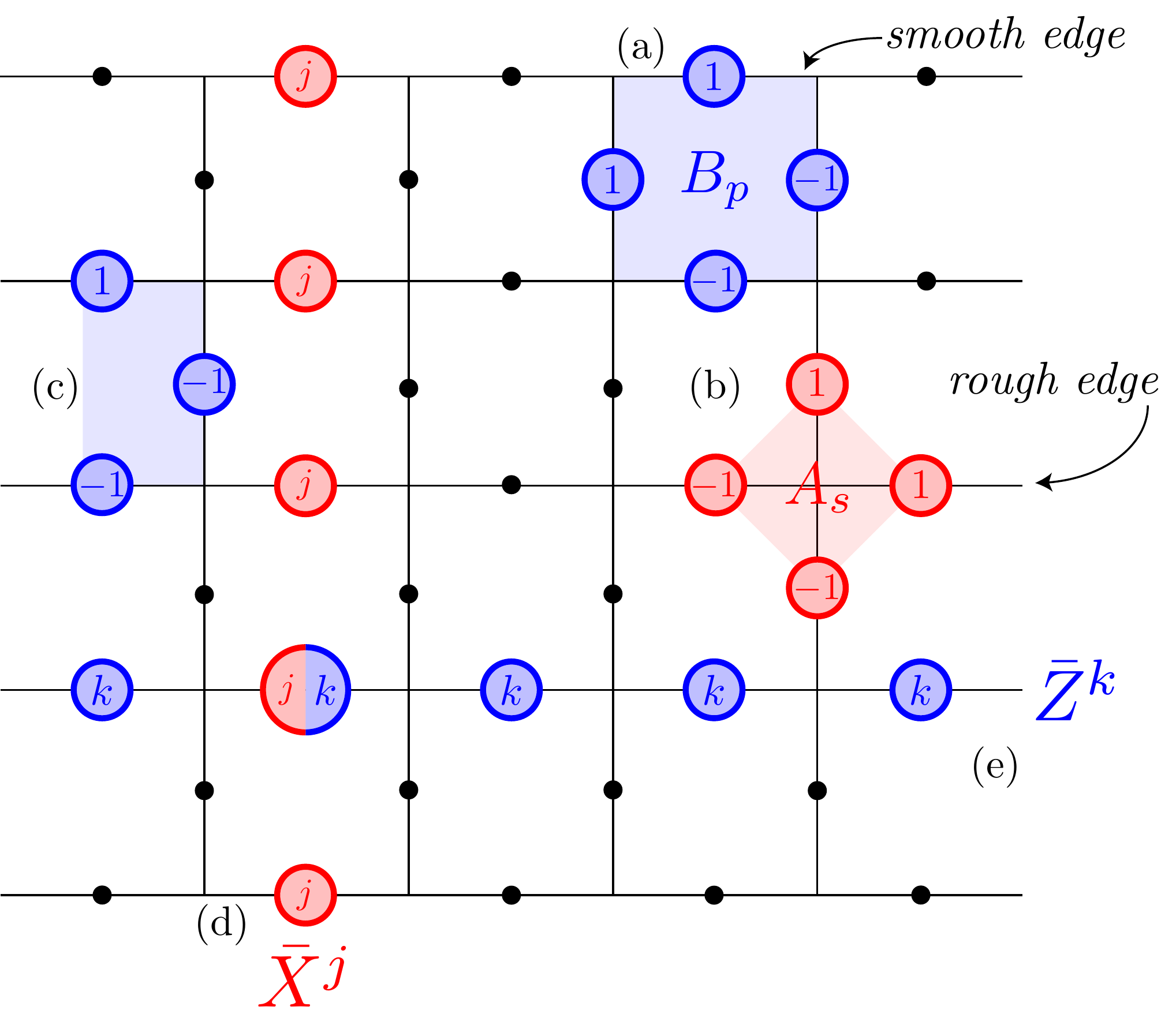} \centering
\caption{(Color online) An example of a distance 5 surface code. Qudits are shown as black dots, arranged on the edges of a lattice with two types of boundary: rough and smooth. For clarity, when an arbitrary $ X^{j} $ or $ Z^{k} $ Pauli operator acts on a physical qudit, we only include the exponents $ j $ and $ k $ on the edges of the figure. We use red for $ X^j $ errors and vertex operators, and blue for $ Z^k $ errors or plaquette operators. (a) and (b) An example of a single plaquette and vertex operator, respectively. (c) An example of a deformed rough edge plaquette operator (3-body operator). Note that the vertex operators are deformed at smooth edges. (d) and (e) An example of a pair of anti-commuting logical operators.}
\label{fig:lattice} 
\end{figure}

In a surface code qudits are identified with the edges of an $L \times L$ lattice with  boundaries as shown in Fig. \ref{fig:lattice}. The surface code is a stabilizer code with two types of stabilizer generators $\mathcal{S} = \left< A_s, B_p \right>$ defined on the lattice as 
\begin{eqnarray}
A_s &=& X_e\otimes X_{e}^{-1}\otimes X_{e}^{-1} \otimes X_{e} \hspace{3mm} \forall \hspace{1mm}e \in V,\\
B_p &=& Z_e\otimes Z_{e}\otimes Z_{e}^{-1} \otimes Z_{e}^{-1} \hspace{5mm} \forall \hspace{1mm}e \in P,
\label{eqn:bulk_stabilizers}
\end{eqnarray}
where $e \in V$ are the edges surrounding a vertex $V$ of the lattice and $e \in P$ are the edges surrounding a plaquette $P$. We refer to $A_s$ as the \textit{vertex} operators, and to $B_p$ as the \textit{plaquette} operators. An example of each is shown in Fig. \ref{fig:lattice}(a) and (b). Note that the two boundary different types (`rough' and `smooth') of the lattice lead to deformations of plaquette and vertex operators at the boundary, respectively.

The surface code supports one logical qudit. The logical operators for the qudit are defined by string-like $X$ (or $Z$) operators. The logical $\bar{X}$ operators connect the two opposing smooth edges, whereas the logical $\bar{Z}$ operators connect the two rough edges. An example of each is shown in Fig. \ref{fig:lattice}(d) and (e). These operators, together with the stabilizer group, generate the group of Pauli operators which map the code space to itself. We denote this group of logical operators $\mathcal{L}$.   We denote the set of logical operators which do not leave the code space invariant as $\mathcal{L}- \mathcal{S}$. 
The distance of a topological code corresponds to the length of shortest possible logical operator, i.e. the distance is $L$.

\begin{figure}
\includegraphics[width=0.35\textwidth]{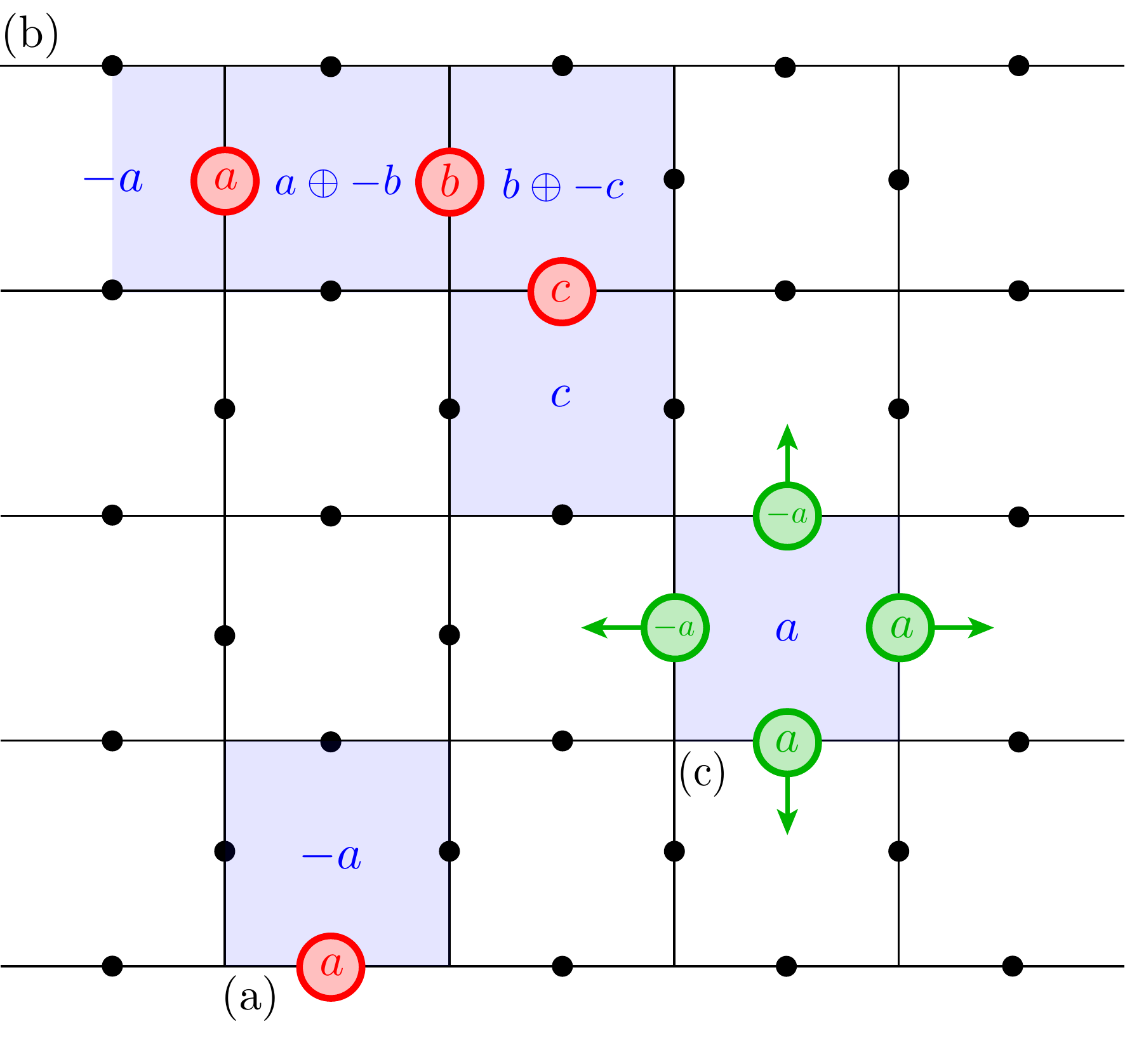} \centering
\caption{(Color online) Examples of $ X $-type errors and the syndrome transportation rule. (a) A single boundary error is only detected by one plaquette. (b) An arbitrary string of three errors and the corresponding intermediate plaquette measurement outcomes.  (c) An example of how to transport the plaquette with outcome `$a$' in any of the indicated directions by applying the relevant $ X $-type operator shown in green.}
\label{fig:errors} 
\end{figure}

Errors that occur on the qudits are detected by measuring the neighbouring stabilizers, with $X$-type and $ Z $-type errors detected independently by the  plaquette and the vertex operators, respectively. This allows us to restrict the discussion to $X$-type errors since results for $Z$-type errors will be analogous. A single $ X $-type error is detected by two adjacent  plaquettes, except when it occurs on a smooth boundary, see Fig. \ref{fig:errors}(a). In general, a string of $ X $-type errors is detected by plaquettes contiguously along the path of the string, as shown by the example in Fig. \ref{fig:errors}(b). This is in contrast to the qubit case ($ d=2 $) where only the end-points of the string give rise to non-trivial plaquette measurements, a situation that can also arise for general $ d $ when the errors along the string possess identical errors. This observation suggests that in higher $d$ the syndrome reveals more information about the path of the errors on the lattice. Indeed it is this information that, if exploited correctly by the decoder, can lead to improved error correction performance, as shown by their higher threshold values, as $ d $ increases. 

We also introduce the concept of syndrome \textit{transportation}: a syndrome can be transported in any direction by applying the appropriate operator as illustrated by the example in Fig. \ref{fig:errors}(c). Moreover, by transporting one syndrome to the location of a second, they are \textit{fused} into a single syndrome such that their charges are added (modulo $d$). These concepts will be useful in Sec. \ref{SecHDRG} when describing our decoder.

Generally speaking, the aim of the decoder is to use the information given by the syndrome to return a correction operator that restores the code to its original state. More formally, let us denote an arbitrary configuration of $ X $-type errors on the 2D surface code by the set $ \textbf{e} $, and the corresponding plaquette measurement outcomes by the set $ \textbf{s}=\{s_{x,y}\} $, where $ s_{x,y}\in \mathbb{Z}_d $ is the outcome of the measurement and the subscripts $x,y$ are the coordinates of the plaquettes, so that $ 1\le x \le L $ and $ 1\le y \le (L-1) $. We will often refer to the outcome $ s_{x,y} $ as the \textit{charge} of the measurement. Then we say that a decoder $ \mathcal{D} $ takes in the syndrome $\textbf{s}$ and returns a correction configuration $ \textbf{f} $. We denote this map by $ \mathcal{D}( \textbf{s}) \rightarrow \textbf{f} $. The decoder succeeds if $ \textbf{e}\otimes \textbf{f}\in \mathcal{S} $ and fails if  $ \textbf{e}\otimes \textbf{f}\in \mathcal{L} - \mathcal{S}$.

In the next section we will give a formal description of the noise model and describe the method for the fault-tolerant simulation.

\section{The Noise Model and Simulation Methods}\label{SecNMSM}

In the literature it is common to test fault-tolerant decoders with a simple error model described by a single parameter $p$. For ease of comparison, we shall follow this convention and use the same error model here. Although this model is not likely to be particularly close to the noise which occurs in physical systems, it has the advantage that it allows $X$- and $Z$-type errors and their correction to be modelled independently. It is thus the standard noise model used to benchmark new decoders.

Between each round of syndrome measurements we assume that each physical qudit is independently subject to an error channel which applies error operator $ X^k $ such that  $ 1\le k\le (d-1) $  with equal probability $ p/(d-1) $, followed by an error channel which applies error operator $ Z^k $ such that  $ 1\le k\le (d-1) $  with equal probability $ p/(d-1) $. We then assume that the outcome of each syndrome measurement $j$ undergoes an error which maps $j$ to $j\oplus k$ for $ 1\le k\le (d-1) $  with equal probability $ p/(d-1) $. Since $X$-type errors, $Z$-type errors and measurement errors are uncorrelated this is often called the \textit{uncorrelated noise model}. 

We estimate the threshold via a Monte Carlo simulation. We shall study a distance $L$ code for a variety of values of $L$. This corresponds to an $L\times L$ surface code grid. For simplicity, we shall let the number of time-steps in our simulation also equal $L$. 

The simulation proceeds by first generating a 3D data structure of $L$ time steps of the accumulated history of the physical qudit errors and the measurement errors. The corresponding syndrome measurement outcomes, taking into account both of these error sources, are then computed.

In order to achieve the close analogy for the relationship between errors and syndromes in the 2D measurement-error-free and 3D general case, Dennis et al \cite{DKLP02} showed that it is most convenient to represent the history of the syndrome outcomes as a 3D grid of syndrome \textit{changes}. 

Let us denote $\textbf{s}_t$ as the set of syndrome outcomes at the $t$-th time step. The set of syndrome changes $\textbf{s}'_t$ at time step $t$ is then defined as the elementwise difference, modulo $d$, of $\textbf{s}_t$ and $\textbf{s}_{t-1}$, i.e. $\textbf{s}'_t=\textbf{s}_t \ominus \textbf{s}_{t-1}$, where `$ \ominus $' denotes subtraction modulo $ d $ and we assume that $\textbf{s}'_1=\textbf{s}_1$. Each set of syndrome changes corresponds to a 2D grid of integers, and we combine these grids into a 3D cubic structure with $t=1$ at the bottom and $t=L$ at the top. We call this grid the syndrome changes history and denote it $\textbf{S}^\prime$. It is convenient to introduce a cartesian coordinate system to refer to the elements of $\textbf{S}^\prime$, i.e. $ s_{t,x,y}$ corresponds to the syndrome change at grid point $(x,y)$ at time step $t$.

The input to the decoder is the 3D syndrome \textit{changes} history ${\textbf{S}^\prime}=\{\boldsymbol{s}_{1},\boldsymbol{s}_{2}\ominus\boldsymbol{s}_{1},\dots,\boldsymbol{s}_{L}\ominus\boldsymbol{s}_{L-1}\}$. The decoder takes the syndrome changes history and returns a  3D correction operator $\textbf{F} = \{\textbf{f}_1, \textbf{f}_2, \dots, \textbf{f}_L\}$. To convert this to a physical correction operator that can be applied in 2D we ignore time-like edges and combine the two-dimensional layers corresponding to each time step, to form a 2D correction operator $ \tilde{\textbf{f}} $ that corrects the accumulated errors at the last time step of the surface code. 

In other words, the resultant correction operator, $ \tilde{\textbf{f}} $, is the sum (modulo $d$) of the correction at each qudit location at each time step, i.e. $\tilde{\textbf{f}} = \otimes_{t} \textbf{f}_t$. We say the decoder has succeeded when the product of the accummulated errors on the qudits, and the returned correction operator is within the stabilizer of the code.

If we are operating below threshold then following the 3D decoding we expect almost all of the errors to have been corrected. There is a finite probability however that some small number of errors will remain after the fault-tolerant decoding has been performed. In a realistic setting the error correction would proceed in this way, eliminating all but a small number of errors in each block of $L$ time steps. At the point when the state is read out, these small errors can be accounted for by taking a majority vote on the measurements of the logical operators. 

For the purposes of the simulation however, we need to determine whether the fault-tolerant decoder has introduced a logical error. The conventional way to overcome this problem is to perform an additional round of error correction in 2D with noise-free syndrome measurements, after which we can be certain that all the errors are corrected and a parity check will reveal whether any logical errors have been introduced.

\section{HDRG Decoder with Noisy Syndromes}\label{SecHDRG}

The HDRG decoder has a simple motivation behind its construction: when the error rate is sufficiently low we expect any errors arising on the surface code lattice to be sparse. This in turn means that syndromes are likely to occur in small, well-separated clusters. The HDRG decoder aims to identify clusters of syndromes generated by such local errors and correct them locally within each cluster. If these clusters have been correctly identified, and the clusters are each small enough that they do not span the lattice, then this strategy results in the decoder computing a correction operator that will correct all errors with high probability. In this section we shall give a formal definition of these concepts in the fault-tolerant setting.

\subsection{Decoder Construction}\label{sec:HDRGconstruction}

\begin{figure}
\includegraphics[width=0.47\textwidth]{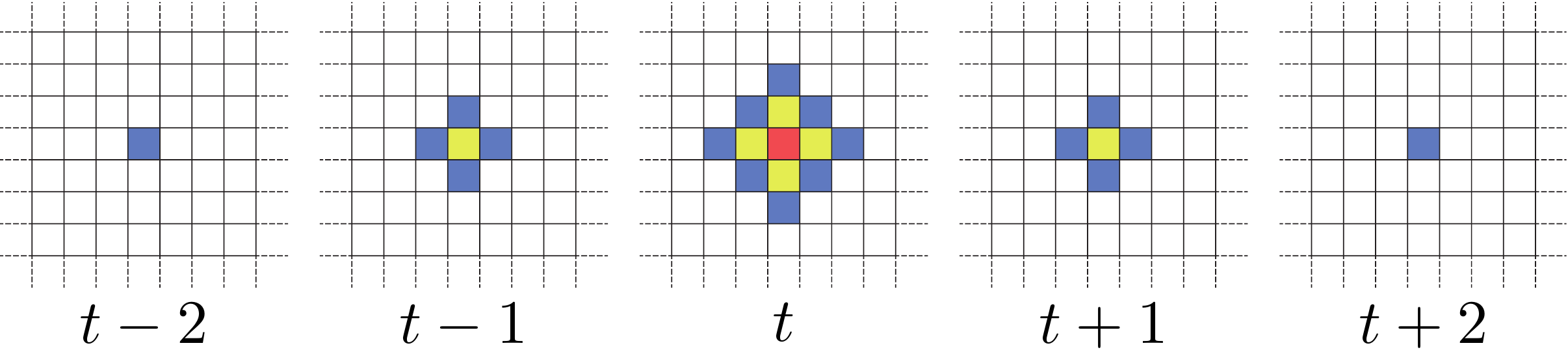} \centering
\caption{(Color online) An illustration of the Manhattan distance metric. The figure shows five time steps from the syndrome changes history. The green (light gray) plaquettes are 1-connected to the central red plaquette  (central plaquette at time step $ t $, medium gray). Blue (dark gray) and green plaquettes are 2-connected to the central plaquette.}
\label{fig:metric} 
\end{figure}

The main concept required for the description of the HDRG decoder is that of a \textit{metric}---a geometric distance function between any pair of elements of a set. In our case, we wish to associate a metric between pairs of syndromes in the set  $ {\textbf{S}'} $. The metric we use is the Manhattan distance, denoted here by $ \delta $, which maps two syndromes as follows:
\begin{equation}
\delta(s_{t,x,y},s_{t^\prime,x^\prime,y^\prime}) = |t^\prime-t| + |x^\prime-x| +|y^\prime-y|,
\end{equation}  
see Fig. \ref{fig:metric} for an illustration for how the region defined by this metric grows.  

We say that two syndromes are $\delta$-\textit{connected} if the distance between them is less than or equal to $\delta$. For a given metric value $ \delta $, we define a cluster $ \mathcal{C}$ to be the set non-trivial syndromes such that every syndrome within the cluster is $\delta$-connected to at least one other syndrome within that cluster. It is easy to see that for a fixed $ \delta $, the syndrome changes history $ {\textbf{S}'} $ can always be partitioned into a set of disjoint clusters such that ${\textbf{S}'} = \mathcal{C}_{1} \cup \mathcal{C}_{2} \cup \cdots \cup \mathcal{C}_{n}$, for some integer $ n $.

Associated to every cluster is a total charge $\oplus_{\mathcal{C}} \: s_{t,x,y}$, where the summation is performed modulo $d$. If this charge is zero, we call the cluster \textit{neutral}. Such a cluster can be annihilated by fusing all of the syndromes contained in the cluster locally, meaning that the Pauli correction operator will have support only within the cluster. If the charge is non-zero but the cluster is $\delta$-connected to any of the three smooth boundaries (two spatial and one time) then we call the cluster \textit{boundary-neutral}. Clusters of this type can be annihilated by fusing the syndromes locally and then connecting the remaining charge to the boundary it overlaps. 

The HDRG decoder involves multiple levels of decoding to fuse together all the elements in $ {\textbf{S}'} $ and return the resultant correction operator. Every decoding level $\ell$ is associated with a distance determining the connectivity of the disjoint clusters at that level. For the metric we have defined we will use $\delta = 2^{\ell}$ starting with $\ell = 0$. This means that the cluster connectivity increases exponentially as we increase the decoding levels. At each level, only the neutral (and boundary-neutral) clusters are fused, leaving any charged clusters to be combined to form neutral clusters at subsequent levels. 

The decoding procedure can now be summarised as follows, starting with $ \ell=0 $.

\begin{enumerate}
\item \textit{Clustering}: Identify all the disjoint $ \delta $-connected clusters at level $ \ell $.
\item \textit{Neutral annihilation}: Fuse each neutral and boundary-neutral cluster locally and return a correction operator.
\item \textit{Renormalize}: If there are clusters that are not annihilated, then increment $ \ell $ by $ 1 $ and return to step $ 1 $.
\end{enumerate}

The decoder stops when there are no non-trivial syndromes remaining. The crucial feature of this decoder is that part of the total correction operator is fixed after each level of decoding. In classical coding theory, decoding algorithms exhibiting such a feature are referred to as a \textit{hard-decision} decoders. An explicit example for a small lattice simulation is illustrated in App. \ref{app:decoder}. 

\subsection{The Run Time of the HDRG Decoder}

The dominant parts of our decoder algorithm that contribute to the run time complexity are the identification of the $\delta$-connected cluster of syndromes (clustering) and the determination of the Pauli operator that eliminates the syndrome (fusion). We shall look at each of these processes in turn and argue that for lower error rates, we expect a run time scaling of $O(L^3)$ and even in the worse case this scaling will be no greater than $O(L^6)$.

Let us first consider the limit in which error rates are low and the errors are extremely sparse. In the clustering part of the algorithm at a given level $\ell$, the algorithm searches a constant number of plaquettes $O(2^{3 \ell})$ around every non-trivial syndrome. 

In the case of extremely sparse syndromes the total number of syndromes is $O(L^3)$ and the decoder will only need to run at the first level $\ell=1$.  Thus in this limit, the dependence of the run time complexity on $L$ for this part of the algorithm will be $O(L^3)$. 

In the worst case scenario we consider the most pessimistic estimates for the clustering step of the algorithm. In this case the decoder will run the maximum number of levels $\ell = O(\log_2 L )$ levels. There will be $O(L^3)$ syndromes and the dependence of the run time complexity on $L$ for this part of the algorithm will thus be
\begin{eqnarray}
\sum_{\ell=0}^{O(\log_2 L )} 2^{3\ell} L^3 &\sim& 2^{3 \, O(\log_2 L )} L^3, \nonumber \\
&\sim& O(L^6),
\end{eqnarray}
to leading order. 

For the fusion part of the algorithm, the syndromes can all be moved to a single point in the box enclosing the cluster. This will take a time that scales with the size of the enclosing box, and the maximum size of the box scales with $L^3$.  Note that since the time complexity of modular arithmetic is independent of modulus $d$, thus this scaling is independent of $d$. 

\subsection{Thresholds Estimation and Percolation Limitation} \label{sec:HDRGthresholds}

To estimate the threshold we simulate the entire process of generating $L$ time steps of errors and noisy syndromes, followed by decoding the syndromes. The simulation was done for $N = 10^4$ runs, and repeated for a range of lattice sizes $L$ and error rates $p$.

We determine the threshold $p^{(d)}_{\mathrm{th}}$ using a rescaling method \cite{WHP03,WB14}. Selecting data close to the point where the curves of different $L$ cross (for fixed qudit dimension) we perform a fit to a function of the form 
\begin{eqnarray}
P_{\mathrm{succ}}(x) = A+B x +Cx^2 +DL^{-1/\mu},
\label{eqn:fit}
\end{eqnarray} 
where $x=(p-p_{\mathrm{th}})L^{1/\nu}$, and the final term in the sum represents a finite-size correction to the fitting. 

\begin{figure}
\includegraphics[width=0.47\textwidth]{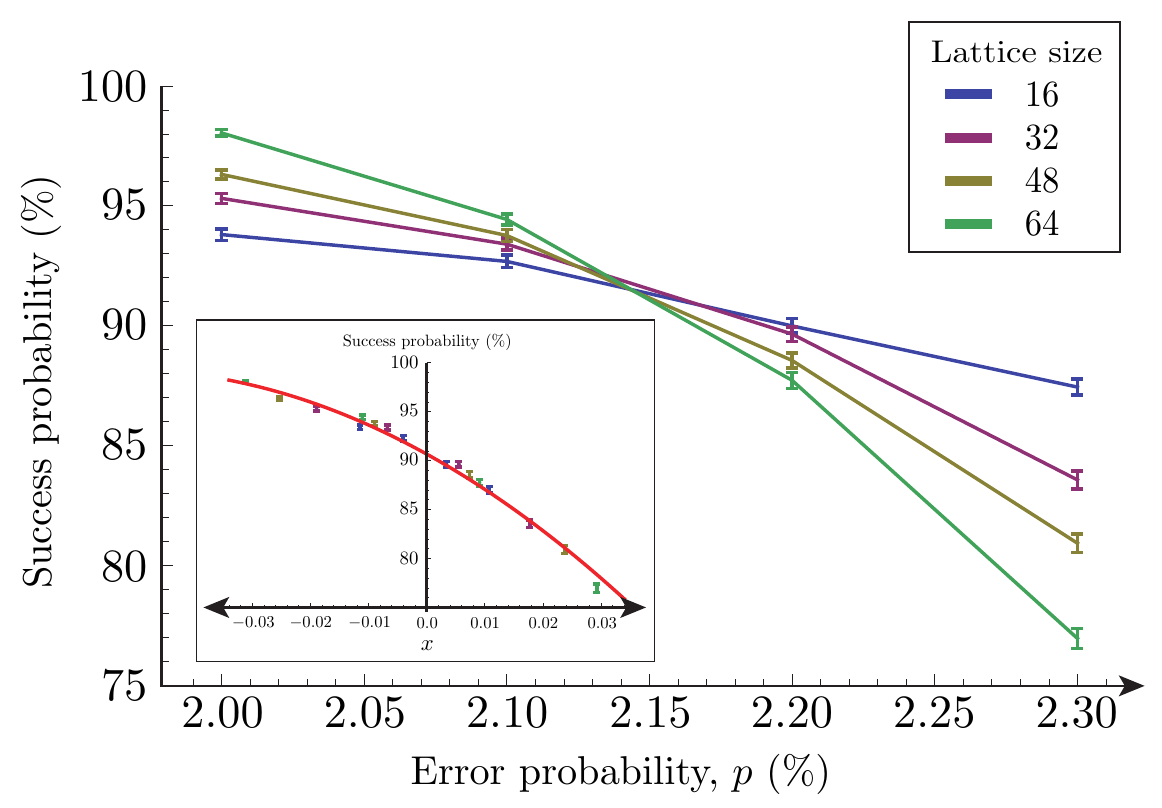} \centering
\caption{(Color online) An example for the collected simulation data used to estimate the threshold for qubits. The inset figure shows the fitting of the function $P_{\mathrm{succ}}(x) = A+B x +Cx^2 +DL^{-1/\mu}$ to the rescaled data.  }
\label{fig:qubit} 
\end{figure}

The success probability of the decoder for the qubit case is shown in Fig. \ref{fig:qubit}, where we find a threshold value of $p_{\mathrm{th}}^{(2)} = 0.0215 \pm 0.0006$. This allows us to directly compare our decoder with other fault-tolerant qubit decoders, for example the soft-decision renormalization group decoder by Duclos-Cianci and Poulin achieves a threshold of $p^{(2)}_{\mathrm{th}}=0.019 \pm 0.004$ \cite{CP14}.

\begin{figure}
\includegraphics[width=0.47\textwidth]{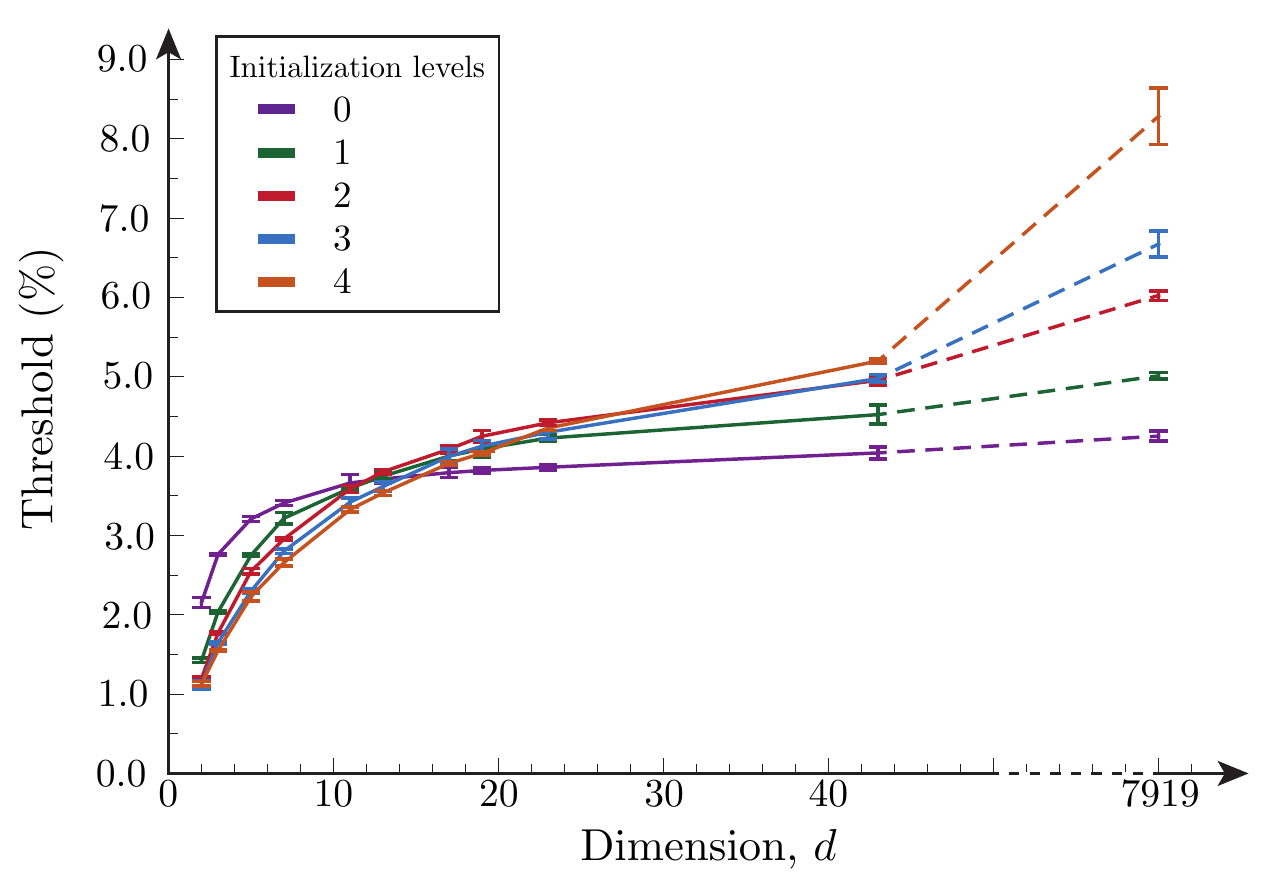} \centering
\caption{(Color online) A summary of all qudit thresholds for different number of rounds of the initialization step. We have chosen the 1000th prime dimension ($d=7919$) to represent the asymptotic limit. Although for small qudit dimensions the initialization step disrupts the syndrome too much and reduces the threshold, we see that in the asymptotic limit there is a clear advantage to using this technique.}
\label{fig:thresholds} 
\end{figure}

Using the same technique of rescaling and fitting the function in Eq. \ref{eqn:fit} we can determine the threshold $p_{\mathrm{th}}^{(d)}$ for further qudit dimensions. Although our HDRG decoder works for arbitrary qudit dimension $d$ we consider the first few prime dimensions, and in order to determine the asymptotic behaviour we also consider one very high qudit dimension, $d=7919$, the 1000th prime number. The results are shown as the plot labelled `Initialization levels 0' in Fig. \ref{fig:thresholds}. The plot shows that the threshold achieved by the decoder increases monotonically with increasing qudit dimension, but quickly saturates to a maximum value of $p_{\mathrm{th}}^{(7919)} = 0.042 \pm 0.09$. Previous work performed on the noiseless syndrome measurement version of the HDRG in \cite{ABCB14} suggests that this saturation is due to a syndrome percolation effect.

\begin{figure}
\includegraphics[width=0.49\textwidth]{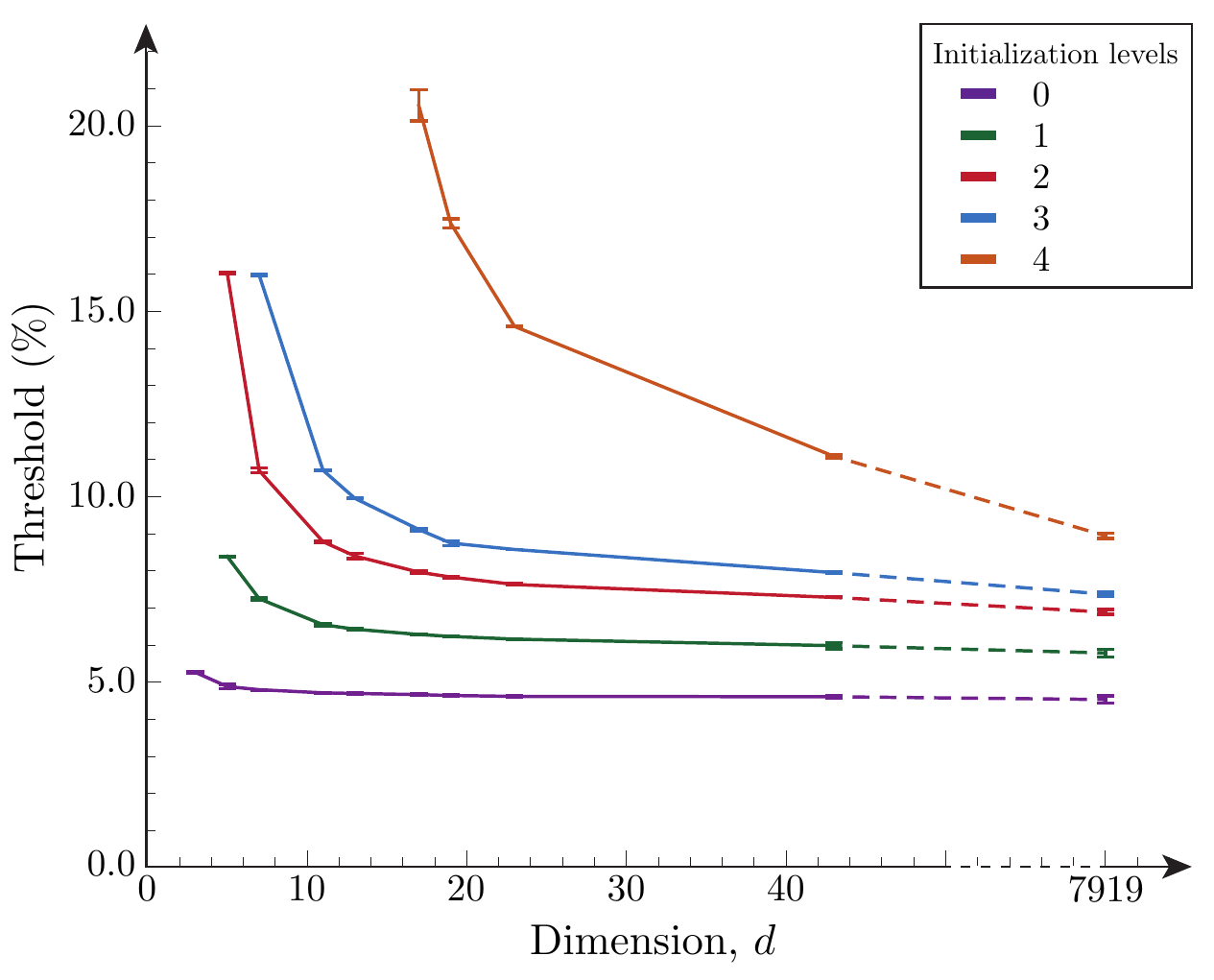} \centering
\caption{(Color online) A summary of percolation thresholds for different number of rounds of the initialization step. The initialization step disrupts the percolation for low qudit dimensions meaning that in some cases a threshold cannot be identified. }
\label{fig:percolation} 
\end{figure}

In order to verify this hypothesis, we performed a simulation of the syndrome percolation threshold. This was done by generating the qudit noise and noisy syndrome measurements for each qudit dimension in the same way as for the decoder simulation. However, once the syndrome changes were calculated, we performed a check to determine whether any $1$-connected clusters in ${\textbf{S}'}$ percolated the lattice in the $x$ or $y$ directions. The $t$ direction was not checked since we want to determine whether the percolating cluster is able to support a logical operator once it is collapsed to $\tilde{\textbf{f}}$, and any string-like operators in the $t$ direction are unphysical. This information is summarised in the plot labelled `Initialization levels 0' in Fig. \ref{fig:percolation}. The saturated value for the percolation thresholds for dimension $7919$ is around $4.5$\%, agreeing with our prediction that the HDRG decoder thresholds are upper-bounded by the percolation threshold.

In the next section we show how to overcome this syndrome percolation effect and achieve improved qudit thresholds using an initialization step. This is an algorithm which is run before the HDRG to disrupt the percolating clusters. 
 
\subsection{Further Enhancement}

The initialization step is a subroutine that sweeps through all of the syndromes ${\textbf{S}'}$ searching for \textit{neutral sub-clusters} in order to disrupt the percolating clusters. Unlike the HDRG algorithm, the initialization step does not divide the observed syndrome into disjoint clusters, but simply identifies and eliminates neutral sub-clusters locally.  

As with the decoder, the initialization step has `levels' defined by a metric. However, sub-clusters are more than $\delta$-connected plaquettes, they are 1-connected \textit{paths} of plaquettes, where the charge of the sub-cluster is counted along the entire path. This is because of the fact illustrated in Fig. \ref{fig:errors}, that a connected path of errors will result in a connected neutral path of syndromes. 

An important idea needed to understand the initialization levels is that of \textit{degeneracy} of paths. If there are $\pm h$ steps in the $x$ direction, $\pm v$ steps in the $y$ direction and $\pm z$ steps in the $t$ direction of the path then its degeneracy is given by 
\begin{eqnarray}
D = \frac{(h + v + z)!}{h! \: v! \: z!}.
\end{eqnarray}

\begin{figure}
\includegraphics[width=0.45\textwidth]{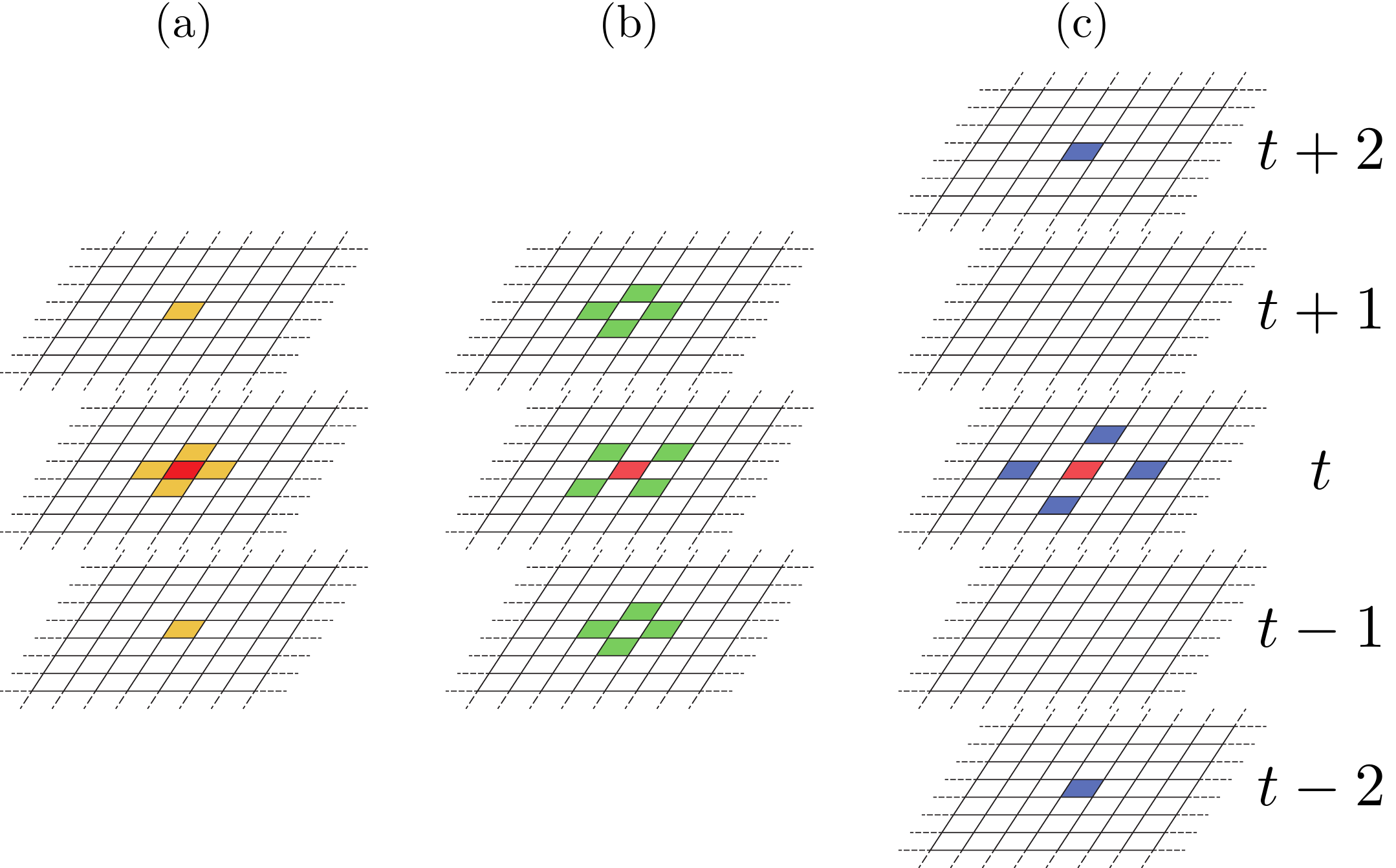} \centering
\caption{(Color online) An illustration of the first three initialization levels. (a) First initialization level. Orange plaquettes (light gray) are 1-connected to the central red plaquette (central plaquette at time step $ t $). (b) Second initialization level. Green plaquettes (medium gray) are a 2-connected to the central plaquette and paths between the central plaquette and any green plaquette have a degeneracy of 2. (c) Third initialization level. Blue plaquettes (dark gray) are a 2-connected to the central plaquette and paths between the central plaquette and any blue plaquette have a degeneracy of 1.}
\label{fig:initial} 
\end{figure}

The initialization levels are defined sequentially by distance from the central syndrome, and the degeneracy of the paths, favouring those paths with equal distance but higher degeneracy as more likely. In Fig. \ref{fig:initial} we show the outer syndromes of the first three initialization levels.

For a given syndrome $ s_{t,x,y} $ we denote by $ \mathcal{Q}={q_1,q_2,\dots,q_n}$ the set of syndromes with the same distance from $ s_{t,x,y} $, and whose paths connecting to $ s_{t,x,y} $ have the same degeneracy. Denote by $ p_{i} $ a possible path connecting $ s_{t,x,y} $ to $ q_i $ and refer to each path as a sub-cluster at the initialization level $k$. 

The \textit{initialization step} $ \mathcal{I}_k $ of \textit{depth} $ k $ consists of running all the levels $ \mathcal{I}_1, \mathcal{I}_2, \dots, \mathcal{I}_k  $ where the initialization procedure $\mathcal{I}_j$ consists of the following steps, beginning with the first non-trivial syndrome:
\begin{enumerate}
\item \textit{Neutral sub-cluster}: search over all the paths $ p_{i} $. If a neutral path (sub-cluster) is identified go to step 2. If all the paths are searched and none of them are neutral, increment the syndrome index by 1 and repeat step 1.
\item \textit{Sub-cluster annihilation}: annihilate the neutral sub-cluster by fusing the syndromes within the sub-cluster i.e. along the path.
\end{enumerate}
We refer to the HDRG decoder when augmented with initialization at a certain depth as the enhanced-HDRG. It is clear that the initialization step is not efficient because the number of paths to search over increases factorially as the depth increases, but for small numbers of levels the number of sub-clusters to search over is still not too high. For example, at the first level of initialization, in the worst case there will be $ 6 $ paths to check for each element in the bulk of $ {\textbf{S}'} $ (corresponding to the 6 neighbouring syndromes, see Fig. \ref{fig:initial}). In general, the initialization step has an overhead of $ C_i L^{3} $ where $C_i$ is the number of paths for each syndrome for the $i$th initialization level. Specifically, $C_i = 6, 24, 6,$ and $ 48 $, for the first four initialization levels respectively. 

We simulated the enhanced-HDRG decoder in the same way described in Sec. \ref{sec:HDRGthresholds}. The results are summarized in Fig. \ref{fig:thresholds}. We see that the asymptotic threshold achieved for four levels of the initialization step is around $8.2$\%. 

Although the improved thresholds for high $d$ suggest that we are successfully able to disrupt the syndrome percolation using this technique, we still observe some saturation of the thresholds. To test this, we performed syndrome percolation simulations using the initialization step prior to the test for percolation. The results are summarized in Fig. \ref{fig:percolation}. We see that the percolation threshold still upper-bounds the enhanced-HDRG thresholds for the corresponding initialization step. 

Despite its success for very high qudit dimensions the enhanced-HDRG is not useful for low qudit dimensions, where the initialization step disrupts the syndromes in a way that results in a \textit{lower} threshold. This can be understood as a result of using a decoding strategy that is too local---the neutral sub-clusters identified are in fact fragments of larger errors and the syndromes do not contain enough information to reconstruct them correctly. This suggests that the syndromes for very high qudit dimensions contain enough information to allow many rounds of initialization to keep improving the threshold. For smaller qudit dimensions however, we see there is an optimal number of initialization rounds that should be performed, for example for $d=17$ we found that the two initialization levels is optimal.

\section{Discussion} \label{SecConclusion}
 
We have presented a modified version of the HDRG decoder that was first introduced by Bravyi and Haah in \cite{BH13} and studied its decoding performance for the surface code with noisy syndrome measurements. The main difference in our version is the use of a more refined metric which has led to an improved threshold. We have chosen the Manhattan distance metric $ \delta $, whereas Bravyi and Haah considered the $ d_\infty $ metric. In our investigations we discovered that the majority of the syndromes are cleared at the first level of decoding. This means that having a more refined metric matters more at $\ell=0$ than it does at higher decoding levels. The $\delta$ metric ensures that the clusters at the first decoding level are as connected as possible by allowing a single syndrome to be connected only to its $6$ nearest neighbour plaquettes. This refinement of the metric is the reason for our improved thresholds.  

We found that, similarly to the measurement noise-free setting, for all but the smallest dimension $d$, syndrome percolation places an upper bound on the decoder threshold for the HDRG decoder. We have demonstrated that this can be overcome by adopting an extra initialization step, which, by scanning for locally neutral sub-clusters, breaks up the percolated lattice allowing the decoder to succeed above the percolation threshold. This has a particularly stark effect for high dimensions, increasing the threshold by almost a factor of two.

The uncorrelated noise model chosen here was adopted for ease of comparison with other decoders. However, an uncorrelated noise model is unlikely to be encountered in experiment. When the dimension is high, in an isotropic depolarising noise model, there would be a high correlation between the presence of $X$-type and $Z$-type errors. A decoder which used this information might achieve significantly higher thresholds. Nevertheless, we expect the decoder presented here to possess an error threshold for any noise model acting independently and identically distributed (i.i.d.) with respect to individual qudits and also non-i.i.d. noise models where the correlation between qudit errors is limited. Testing these possibilities is a pertinent open question.

A remarkable feature of this decoder is the independence of its run time complexity with respect to qudit dimension. This is in stark contrast to other known qudit decoders. For example, the soft-decision renormalization group decoder in the fault-tolerant setting \cite{CP14} has a straightforward implementation in higher dimensions but comes with a cost of a polynomial overhead in $d$ which means its applicability is limited to low dimensions.

To make our comparisons with other decoding algorithms more concrete in the qubit case, further research should focus on a full gate-error simulation of the HDRG in the low noise regime. A comparison  of success probability vs error rate in this regime would allow one to compare overheads, the most relevant figure of merit for judging the relative performance of these decoders.

Given the excellent performance of the MWPMA, the most important applications of the methods here will be for codes where MWPMA is not a suitable decoder. The surface codes studied in this paper are not the only quantum error correcting codes for which HDRG type decoders could be beneficial. An efficient decoding algorithm for the more exotic LDPC code, the 4D hyperbolic code, was introduced by Hastings with similar `greedy local matching' principles as the HDRG decoder \cite{H13}. Other LDPC codes exist for which efficient decoders have not yet been identified \cite{TZ09,KP12}. The development of computationally light fault-tolerant decoders for these codes is essential if they are to be practical. HDRG decoders have demonstrated the efficiency needed to support large scale fault-tolerant error correction on the surface code and a flexibility which may make them well suited to unlock the potential of future novel topological codes. 

\subsection*{Acknowledgements}

We would like to thank Earl Campbell and James Wootton for useful discussions and helpful comments on this manuscript. We acknowledge the Imperial College HPC Service and Legion@UCL for computational resources and associated support services. FHEW and HA acknowledge the financial support of the EPSRC (grant numbers  EP/G037043/1 and EP/K022512/1, respectively). Note: while this work was in preparation, the authors learnt of a similar investigation of the HDRG decoder with non-perfect syndrome measurements (without the initialization step) \cite{HLW14}.

\appendix

\onecolumngrid
\section{Explicit Examples}
In this section we present two explicit examples outlining all the steps of the 3D fault-tolerant simulation for a single sample of errors. The reader may find the figures below more transparent in explaining how the HDRG decoder works in comparison to the description provided in Sec. \ref{SecHDRG}. In both examples we choose a lattice of distance $ 5 $ and qudit dimension $ d=5 $. 

\subsection{Example 1: HDRG Decoding} \label{app:decoder}

In this example we present the simulation for the HDRG decoder without any initialization step. We describe the steps of the simulation in the captions of the following 6 figures. 

\begin{center}

\begin{figure}[h]
\centering
\includegraphics[width=0.9\textwidth]{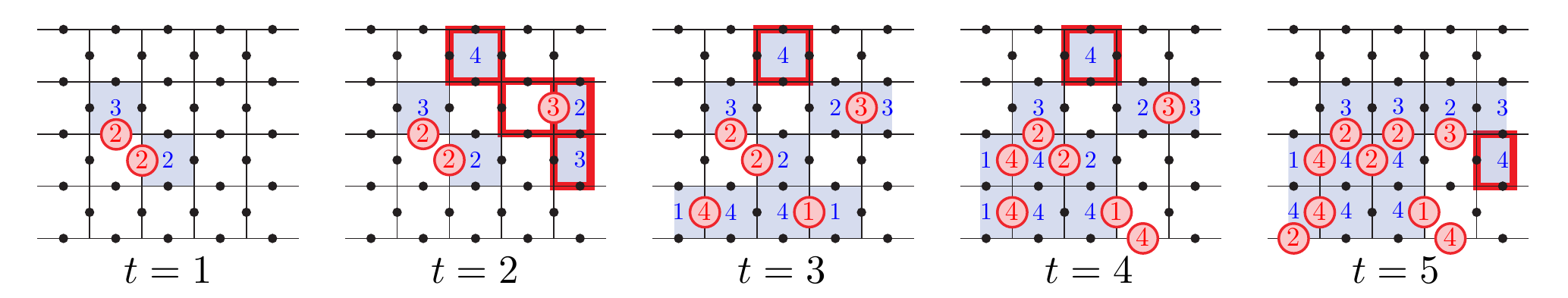}
\vspace{-2mm}
\caption{(Color online) \textbf{Error and Syndrome Histories}: The first step is to generate the full history of errors and noisy syndrome measurements $\textbf{S}$ for $ L=5 $ time steps. The red circles and squares indicate the location of errors. Notice how the errors accumulate at each time step. The goal of the decoder is to correct the final error configuration $ t=5 $.}\label{fig:Appendix1:1}
\end{figure}

\begin{figure}[h!]
\includegraphics[width=0.9\textwidth]{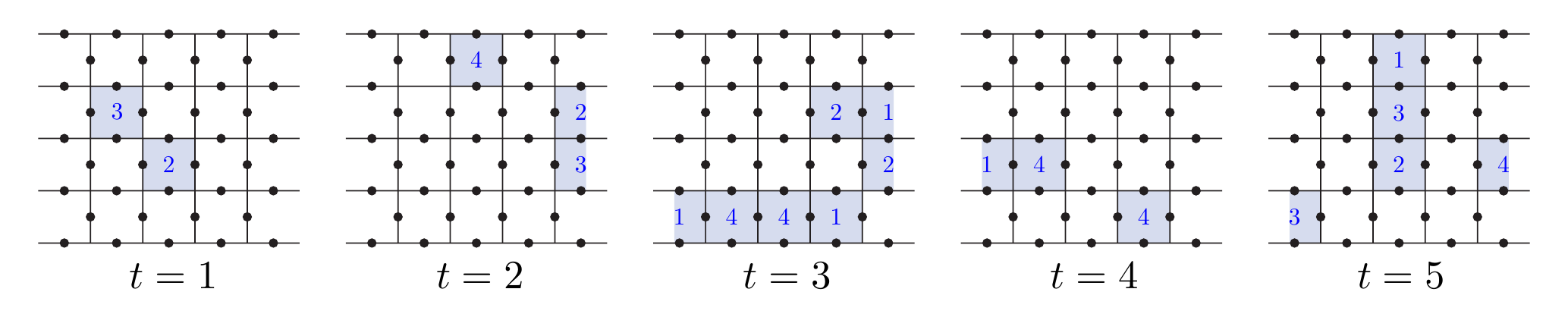}
\vspace{-2mm}
\caption{(Color online) \textbf{Syndrome Changes History}: The second step is to evaluate the syndrome changes history ${\textbf{S}'}=\{\mathbf{e}_1,{\mathbf{e}_2\ominus\mathbf{e}_1}, \mathbf{e}_3\ominus\mathbf{e}_2, \mathbf{e}_4\ominus\mathbf{e}_3, \mathbf{e}_5\ominus\mathbf{e}_4\}$, where the subtraction is performed modulo $ d $. The changes history is passed to the decoder which must infer a correction operator from the information in $ {\textbf{S}'} $ alone.}\label{fig:Appendix1:2}
\end{figure}

\begin{figure}[h!]
\includegraphics[width=0.9\textwidth]{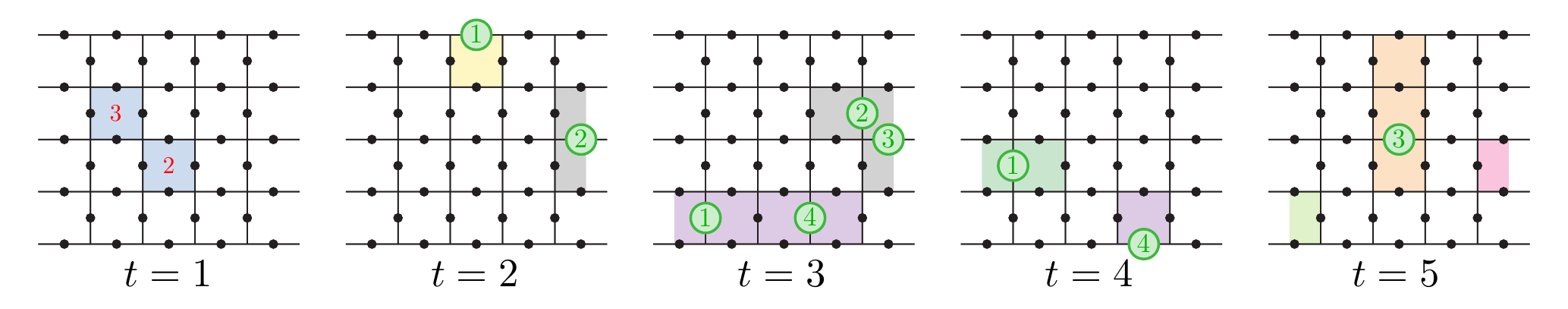}
\vspace{-2mm}
\caption{(Color online) \textbf{HDRG $ \ell=1 $}: The first level of the HDRG decoder divides the set $ {\textbf{S}'} $ into disjoint 1-connected clusters (i.e. $ \delta=1 $). There are three different types of clusters shown: \textit{non-neutral}, \textit{neutral}, and \textit{boundary-neutral}. Specifically, there are two single element non-neutral clusters shown in blue with the their charge displayed. These clusters cannot be fused at this level. Moreover, there are two neutral clusters in the bulk (grey and dark green), meaning that their total charge adds to zero (modulo $ 5 $). The elements of each neutral cluster are fused together to the vacuum. Finally, there are five boundary-neutral clusters (yellow, purple, light green, orange and pink). The total charge of these clusters do not add up to zero, but since they are 1-connected to one of the boundaries they can be fused with that boundary. When the cluster is fused with the time boundary, no physical correction is applied. The resultant correction operator from the fusion of the neutral and boundary-neutral clusters is shown in green.}
\end{figure}

\begin{figure}[h!]
\includegraphics[width=0.9\textwidth]{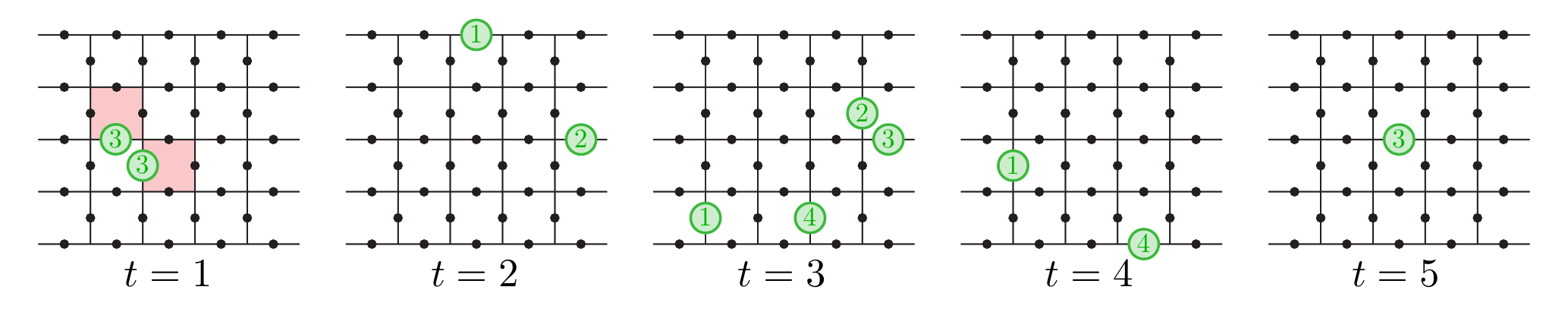}
\vspace{-2mm}
\caption{(Color online) \textbf{HDRG $ \ell=2 $}: The only remaining non-neutral cluster from the previous level is now 2-connected (shown in red). Its elements are fused together and a local correction is returned.}\label{fig:Appendix1:4}
\vspace{1cm}
\includegraphics[width=0.9\textwidth]{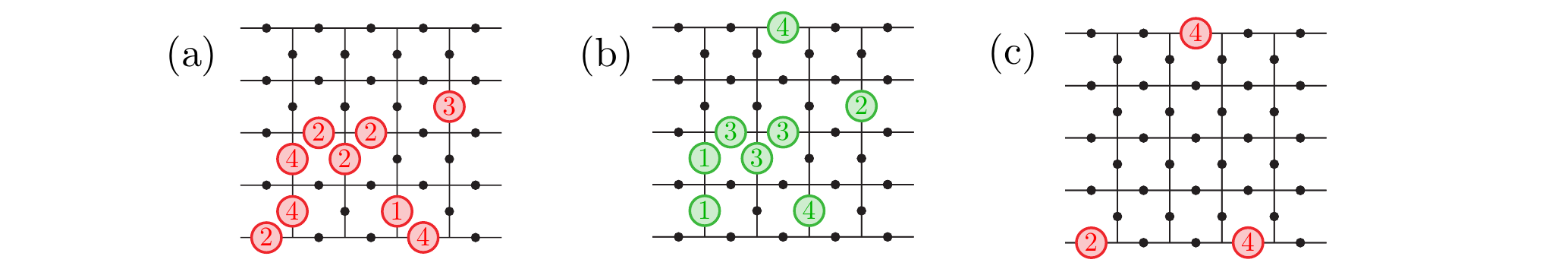}
\vspace{-2mm}
\caption{(Color online) \textbf{Projected Correction}: (a) The final (physical) error layer at $t=5$. (b) Projected correction operator from corrections identified in Fig. \ref{fig:Appendix1:4}, $\tilde{\textbf{f}} = \textbf{f}_1 \otimes \textbf{f}_2 \otimes \textbf{f}_3 \otimes \textbf{f}_4 \otimes \textbf{f}_5 $, which is equivalent to summing the exponents of the operators modulo $d$. (c) The product of the accumulated error and the projected correction operators. The correction has resulted in a small number of remaining errors. }
\vspace{1cm}
\includegraphics[width=0.9\textwidth]{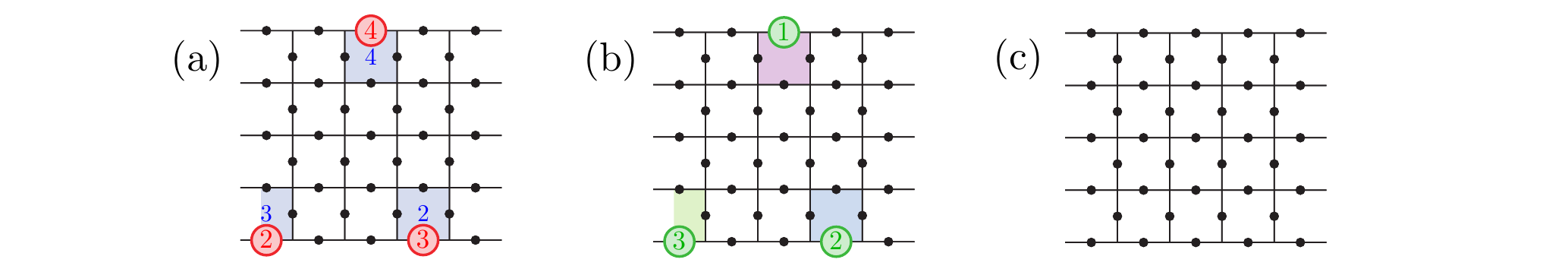}
\vspace{-2mm}
\caption{(Color online) \textbf{Noise-Free Decoding}: To confirm whether the decoder has succeeded or failed, we must perform an additional round of decoding with noise-free syndrome measurements. (a) The outcomes of the noise-free syndrome measurements. (b) Clustering and correction operators. (c) Result of noise-free decoding. As we can see in this instance all the errors have been eliminated, no logical error has been introduced and the decoding has succeeded.}
\label{fig:appendix:6}
\end{figure}

\end{center}

\clearpage
\newpage
\subsection{Example 2: Enhanced-HDRG Decoding with a Depth 1 Initialization Step}

In this example we present the simulation for the HDRG decoder when augmented with the first level of initialization, $ \mathcal{I}_1 $. The initialization step is shown in Fig. \ref{fig:Appendix2:2}, and all the remaining steps are similar to those shown in the previous example. 

\begin{center}
\begin{figure}[h!]
\centering
\includegraphics[width=0.9\textwidth]{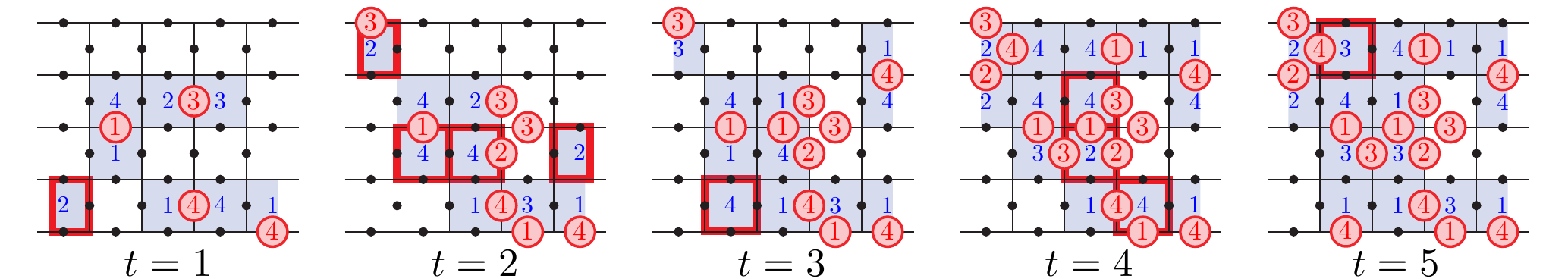}
\includegraphics[width=0.9\textwidth]{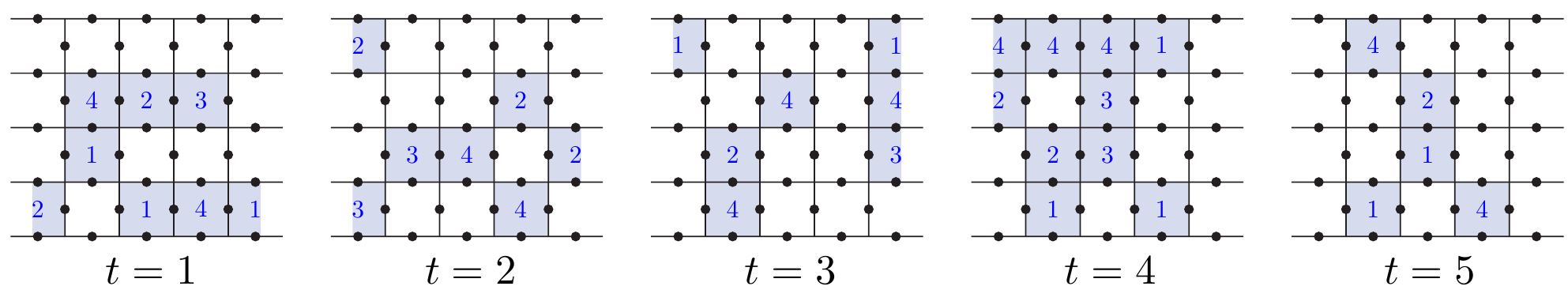}
\caption{(Color online) \textbf{First row}: The generation of the error and syndrome histories, similar to Fig. \ref{fig:Appendix1:1}. \textbf{Second row}: The syndrome changes history $ {\textbf{S}'} $, similar to Fig. \ref{fig:Appendix1:2}. Notice how $ {\textbf{S}'} $ contains a percolating cluster of syndromes.}\label{Fig:Appendix2:1}
\end{figure}

\begin{figure}[h!]
\includegraphics[width=0.9\textwidth]{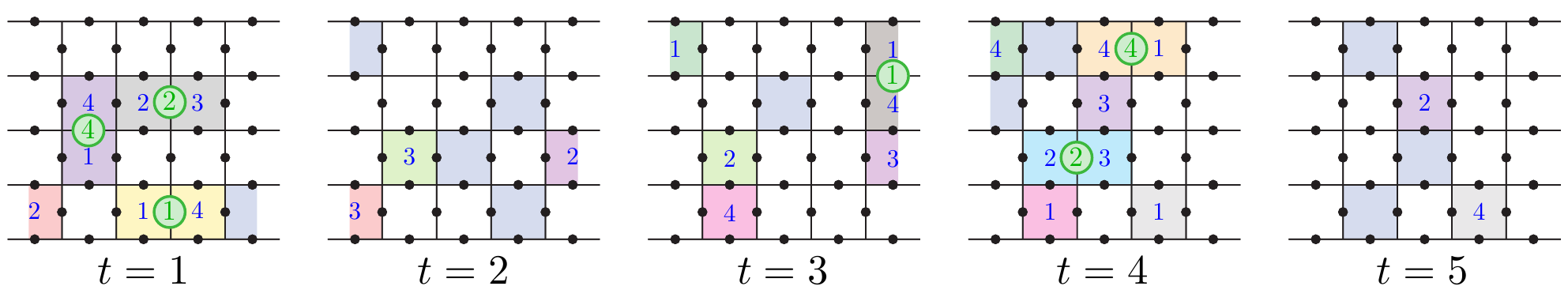}
\caption{(Color online) \textbf{Initialization $ \mathcal{I}_1 $}: At the first level of initialization there are only $ 6 $ sub-clusters to search around each plaquette. The search works by searching over every non-trivial syndrome and checking its $ 6 $ neighbouring plaquettes sequentially to see if any of them form a two-element neutral sub-cluster. Once a neutral pairing is found, the two plaquettes are fused together and a single correction operator is returned. Note that in this step we do not pair plaquettes to the physical or the time boundary. In the figure each sub-cluster is coloured differently. }
\label{fig:Appendix2:2}
\end{figure}

\begin{figure}[h!]
\includegraphics[width=0.9\textwidth]{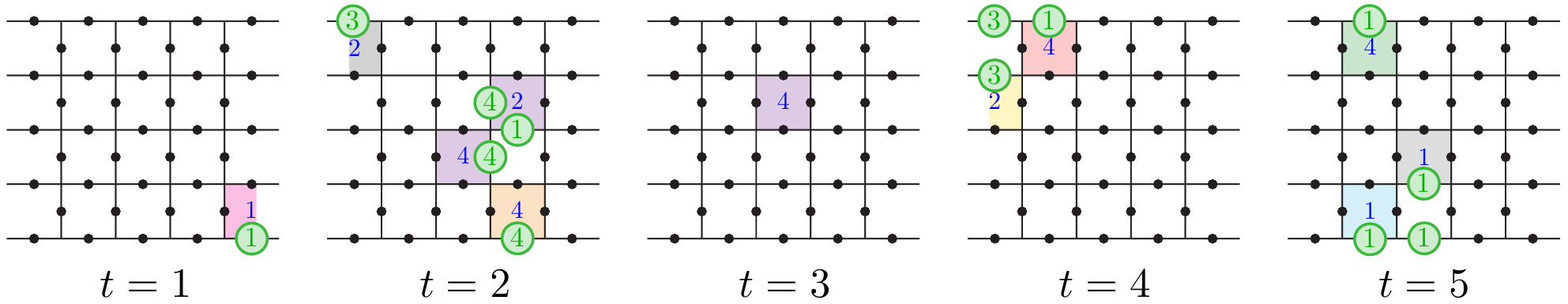}
\caption{(Color online) The initialization step has disrupted the percolating cluster. \textbf{HDRG Decoding}: Neutral and boundary-neutral clusters identified by running two levels of the HDRG decoder with the correction operator returned shown in green.}\label{fig:Appendix2:3}
\end{figure}

\begin{figure}[h!]
\includegraphics[width=0.9\textwidth]{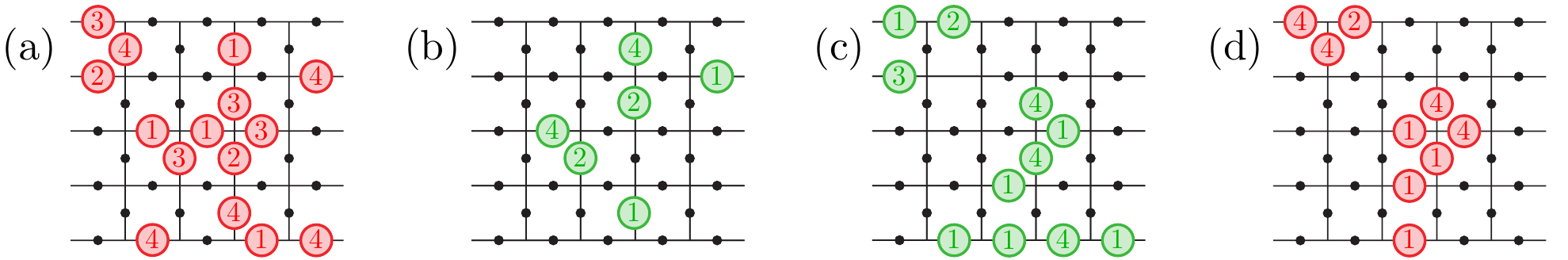}
\caption{(Color online) \textbf{Projected Correction}: (a) The accumulated layer of errors at $ t=5 $, see the top row of Fig. \ref{Fig:Appendix2:1}. (b) The projected correction operator from running the initialization $ \mathcal{I}_1 $, see Fig. \ref{fig:Appendix2:2}. (c) The projected correction operator obtained from the HDRG decoder, Fig. \ref{fig:Appendix2:3}. (d) The resultant errors after taking the operator product of the two correction layers and the accumulated layer of errors.}
\vspace{1cm}
\includegraphics[width=0.9\textwidth]{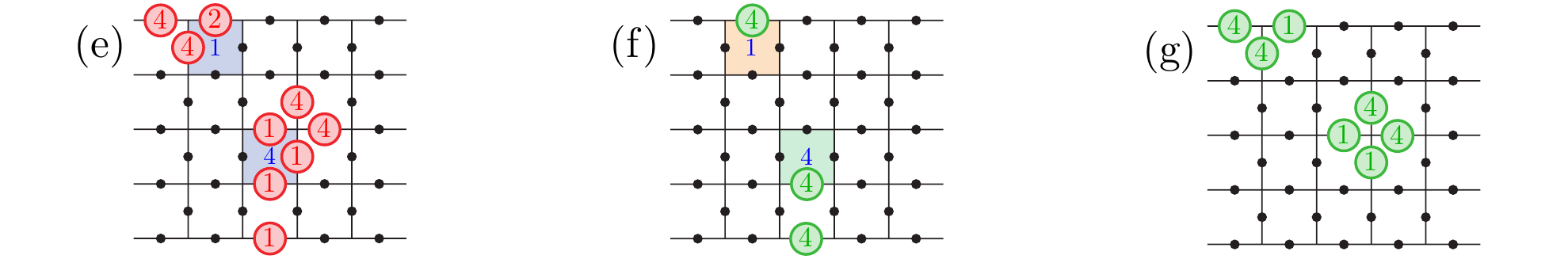}
\caption{(Color online) \textbf{Noise-free Decoding}: (e) Noise-free syndrome measurements. (f) Clustering and correction operators. (g) The result of noise-free decoding. In this case the resultant operators are all members of the stabilizer group so once again the decoding has been successful.}
\end{figure}

\end{center}

\clearpage

\twocolumngrid

% % % % % % %Bibliography % % % % %

% % % % % % % %Appendix % % % % % %
\clearpage

%%%%%%%%%%%%%%%%%%%%%%%%%%%%%%%%%%%%


\begin{thebibliography}{80}

\bibitem{RHG07} R. Raussendorf, J. Harrington, and K. Goyal, New J. Phys. \textbf{9}, 199 (2007).
\bibitem{RH07} R. Raussendorf and J. Harrington, Phys. Rev. Lett. \textbf{98}, 190504 (2007).
\bibitem{DFSG09} S.J. Devitt, A.G. Fowler, A.M. Stephens, A.D. Greentree, L.C.L. Hollenberg, W.J. Munro, and K. Nemoto, New J. Phys. \textbf{11}, 083032 (2009).
\bibitem{FMMC12} A.G. Fowler, M. Mariantoni, J.M. Martinis, and A.N. Cleland, Phys. Rev. A \textbf{86}, 032324 (2012).
\bibitem{BK98} S.B. Bravyi and A.Y. Kitaev, arXiv:quant-ph/9811052.
\bibitem{K03} A.Y. Kitaev, Ann. Phys. \textbf{303}, 2 (2003).
\bibitem{DKLP02} E. Dennis, A. Kitaev, A. Landahl, and J. Preskill, J. Math. Phys. \textbf{43}, 4452 (2002).
\bibitem{WFH11} D.S. Wang, A.G. Fowler, and L.C.L. Hollenberg, Phys. Rev. A \textbf{83}, 020302 (2011).
\bibitem{S14} A.M. Stephens, Phys. Rev. A \textbf{89}, 022321 (2014).
\bibitem{BK05} S. Bravyi and A. Kitaev, Phys. Rev. A \textbf{71}, 022316 (2005).
\bibitem{ACB12} H. Anwar, E.T. Campbell, and D.E. Browne, New J. Phys. \textbf{14}, 063006 (2012).
\bibitem{CAB12} E.T. Campbell, H. Anwar, and D.E. Browne, Phys. Rev. X \textbf{2}, 041021 (2012).
\bibitem{C14} E.T. Campbell, arXiv:1406.3055.
\bibitem{CP13} G. Duclos-Cianci and D. Poulin, Phys. Rev. A \textbf{87}, 062338 (2013).
\bibitem{ABCB14} H. Anwar, B.J. Brown, E.T. Campbell, and D.E. Browne, New J. Phys. \textbf{16}, 063038 (2014).
\bibitem{WAK14} J.R. Wootton, R.S. Andrist and H.G. Katzgraber, arXiv:1406.5974.
\bibitem{SA13} A Smith, B.E. Anderson, H. Sosa-Martinez, C.A. Riofr\'io, I.H. Deutsch, and P.S. Jessen, Phys. Rev. Lett. \textbf{111}, 170502 (2013).
\bibitem{ASM14} B.E. Anderson, H. Sosa-Martinez, C.A. Riofr\'io, I.H. Deutsch, and P.S. Jessen, arXiv:1410.3891.
\vfill\eject
\bibitem{WHP03} C. Wang, J. Harrington, and J. Preskill, Ann. Phys. \textbf{303}, 31 (2003).
\bibitem{FWH12} A.G. Fowler, A.C. Whiteside and  L.C.L. Hollenberg, Phys. Rev. Lett. \textbf{108}, 180501 (2012).
\bibitem{F15} A.G. Fowler, Quant. Inf. and Comp. \textbf{15}, 0145 (2015).
%\bibitem{FWHPRA12} A.G. Fowler, A.C. Whiteside and  L.C.L. Hollenberg, Phys. Rev. A \textbf{86}, 042313 (2012).
\bibitem{CP14} G. Duclos-Cianci and D. Poulin, Quant. Inf. Comp. \textbf{14}, 0721 (2014).
\bibitem{HLW14} A. Hutter, D. Loss, and J.R. Wootton, arXiv:1410.4478.
\bibitem{HWL14} A. Hutter, J.R. Wootton, and  D. Loss, Phys. Rev. A \textbf{89}, 022326 (2014).
\bibitem{HCEK14} M. Herold, E.T. Campbell, J. Eisert, and M.J. Kastoryano, arXiv:1406.2338.
\bibitem{BSV14} S. Bravyi, M. Suchara, and A. Vargo, Phys. Rev. A \textbf{90}, 032326 (2014).
\bibitem{W13} J.R. Wootton, arXiv:1310.2393.
\bibitem{BH13} S. Bravyi and J. Haah, Phys. Rev. Lett. \textbf{111}, 200501 (2013).
\bibitem{OAIM04} T. Ohno, G. Arakawa, I. Ichinose, and T. Matsui, Nucl. Phys. B \textbf{697}, 462 (2004).
\bibitem{H04} J.W. Harrington, PhD thesis, California Institute of Technology, (2004).
\bibitem{K02} A. Kitaev, Proc. Sym. App. Math. \textbf{58}, 267 (2002).
\bibitem{BB07} S.S. Bullock and G.K. Brennen, J. Phys. A \textbf{40}, 3481 (2007).
\bibitem{FWH10} A.G. Fowler, D.S. Wang, and L.C.L. Hollenberg, Quant. Info. Comput. \textbf{11}, 1\&2 (2011).
\bibitem{WB14} F.H.E. Watson and S.D. Barrett,  New J. Phys. \textbf{16}, 093045 (2014). 
\bibitem{H13} M.B. Hastings, arXiv:1312.2546.
\bibitem{TZ09} J.-P. Tillich, G. Zemor, IEEE Int. Symp. Info. Theo., 799 (2009).
\bibitem{KP12} A.A. Kovalev and L.P. Pryadko, IEEE Int. Symp. Info. Theo., 348 (2012).


\end{thebibliography}
\end{document}